\documentclass[a4paper,12pt]{spieman}  
\usepackage{amsmath,amsfonts,amssymb}
\usepackage{graphicx}
\usepackage{setspace}
\usepackage{tocloft}

\usepackage{siunitx} 
\usepackage{multirow}

\title{Male pelvic synthetic CT generation from T1-weighted MRI using 2D and 3D convolutional neural networks}

\author[1,2]{Jie Fu}
\author[2]{Yingli Yang}
\author[1,2]{Kamal Singhrao}
\author[2]{Dan Ruan}
\author[2]{Daniel A. Low}
\author[2,*]{John H. Lewis}

\affil[1]{University of California, Los Angeles, David Geffen School of Medicine, 10833 Le Conte Ave, Los Angeles, CA 90095}
\affil[2]{University of California, Los Angeles, Department of Radiation Oncology, 200 Suite B265, Medical Plaza Driveway, Los Angeles, CA 90095}

\cftpagenumbersoff{figure}
\cftpagenumbersoff{table} 
\begin{document} 
\maketitle

\begin{spacing}{1.25}
\begin{abstract} \\
\textbf{Purpose:} The improved soft tissue contrast of magnetic resonance imaging (MRI) compared to computed tomography (CT) makes it a useful imaging modality for radiotherapy treatment planning. Even when MR images are used for treatment planning, standard clinical practice currently also requires a CT for dose calculation and x-ray based patient positioning. This increases workloads, introduces uncertainty due to the required inter-modality image registrations, and involves unnecessary irradiation. While it would be beneficial to use exclusively MR images, a method needs to be employed to estimate a synthetic CT (sCT) for generating electron density maps and patient positioning reference images. We investigated deep learning approaches, 2D and 3D convolutional neural network (CNN) methods, to generate a male pelvic sCT using a T1-weighted MR image and evaluate their performance using geometric and voxel-wise metrics. \\
\textbf{Method:}
A retrospective study was performed using CTs and T1-weighted MR images of 20 prostate cancer patients. The proposed 2D CNN model, which contained 27 convolutional layers, was modified from the SegNet for better performance.  3D version of the CNN model was also developed. Both CNN models were trained from scratch to map intensities of T1-weighted MR images to CT Hounsfield Unit (HU) values. Each sCT was generated in a five-fold-cross-validation framework and compared with the corresponding CT using voxel-wise mean absolute error (MAE). The sCT geometric accuracy was evaluated by comparing bony structures in the CTs and the sCTs using dice similarity coefficient (DSC), recall, and precision. Wilcoxon signed-rank tests were performed to evaluate the differences between the 2D and 3D CNN models.  \\ 
\textbf{Result:}
Generating pelvic sCT datasets required approximately 5.5 s using the proposed deep learning methods. The MAE averaged across all patients were 40.5 $\pm$ 5.4 HU and 37.6 $\pm$ 5.1 HU for the 2D and 3D CNN models, respectively. The DSC, recall, and precision of the bony structures were 0.81 $\pm$ 0.04, 0.85 $\pm$ 0.04, and 0.77 $\pm$ 0.09 for the 2D CNN model, and 0.82 $\pm$ 0.04, 0.84 $\pm$ 0.04, and 0.80 $\pm$ 0.08 for the 3D CNN model, respectively. P values of the Wilcoxon signed-rank tests were less than 0.05 except for recall, which was 0.6. \\
\textbf{Conclusion:}
The 2D and 3D CNN models generated accurate pelvic sCTs for the 20 patients using T1-weighted MR images. The evaluation metrics and statistical tests indicated that the 3D model was able to generate sCTs with better MAE, bone DSC, and bone precision. The accuracy of the dose calculation and patient positioning using generated sCTs will be tested and compared for the two models in the future. 
\end{abstract}
\end{spacing}


{\noindent \footnotesize\textbf{*}John H. Lewis,  \linkable{jhlewis@mednet.ucla.edu} }
\begin{spacing}{1}   

\section{Introduction}
\label{sect:intro}
Magnetic resonance imaging (MRI) is often integrated into radiotherapy treatment planning\cite{RN1}, particularly for tumors in regions like the brain, head and neck, and prostate\cite{RN2}. The superior soft tissue contrast of MR images facilitates precise delineations of tumors and organs at risk\cite{RN3,RN4}. MR images can also provide guidance for adaptive radiation therapy\cite{RN5,RN6}. The standard MRI-guided clinical workflow includes acquisition of a planning computed tomography (CT). The CT Hounsfield Unit (HU) map, essentially a scaled linear attenuation map, is used to generate both digital reconstructed radiographs for subsequent patient positioning and electron density maps for dose calculation. 

The need to acquire the CT when employing MR images for contouring has several disadvantages. Acquiring a CT increases unwanted radiation exposure, clinical workload, and financial cost\cite{RN7}. In addition, co-registering CT and MR images is required for transferring delineation structures from the MR image to the CT. This process introduces a systematic uncertainty, which is estimated to be 2 mm to 5 mm in various sites, that propagates throughout the treatment\cite{RN8}. MR-only radiotherapy can avoid these pitfalls. 

To achieve MR-only radiotherapy, synthetic HU maps, termed synthetic CT (sCT) images, must be accurately generated from the MR images. To date, there are three types of methods developed for this: atlas-based, voxel-based and hybrid\cite{RN8}. In atlas-based methods, a set of one or multiple co-registered MRI-CT images are deformably registered to a patient`s MR image\cite{RN9,RN10,RN11}. The resulting transformation can then be applied on the CT-atlas to generate the sCT. Atlas-based approaches can be time-consuming, particularly when the atlases are large, and often fail if the patient has very different anatomy from what is represented by the atlas. 

Voxel-based methods convert individual MR voxel intensities to HU values using bulk density assignments or machine learning models. Bulk density techniques assign the patient`s electron density either to water or to pre-defined electron densities within selected MR-segmented tissue types.\cite{RN12,RN13,RN14,RN15} These methods may lead to dose discrepancies and often have limited value in generating positioning reference images. Machine learning methods use paired MRI-CT images to train models that associate MRI intensities with HU values. It is challenging for models to distinguish air from bone in conventional MR images as both tissues exhibit weak signals due to their small T2 values. Some learning methods required manual bone segmentation\cite{RN16,RN17} in conventional MR images or require acquisition of specialized MR sequences like ultrashort echo time sequence\cite{RN18,RN19,RN20} for separating bone and air. Some methods used multiple MR images acquired with additional sequences designed to distinguish different tissue types.\cite{RN21,RN22,RN23}. Adding sequences can increase workload and extend scan time. 

Hybrid methods combine elements of voxel-based and atlas-based approaches.\cite{RN11,RN23} A detailed summary of previous approaches can be found in the review paper by Karlsson $et \ al$ {\cite{RN8}}. 

Recently, deep learning models\cite{RN24} proposed to estimate sCTs from MR images have demonstrated promising results. Nie $et \ al.$ \cite{RN25} presented a 3D convolutional neural network (CNN) model with three convolutional layers. It was trained to convert 3D patches of pelvic MR images to corresponding 3D sCT patches. The sCT was then generated by averaging the HU values of overlapping sCT patches. An updated model with an adversarial network\cite{RN26} was later proposed to improve the sCT quality. Training on patches rather than whole volumes reduces the required number of CNN model parameters and saves computational resources. However, using patches might miss larger scale (relative to patch size) image features. A SegNet-like 2D CNN model with 27 convolutional layers was proposed by Han for brain sCT generation. This more complex CNN model could capture long-range information and generate brain sCTs slice by slice without dividing images into patches\cite{RN27,RN28}. 

Ignoring other model- and data-specific variations, it is reasonable to expect that 3D models should have better performance than their corresponding 2D models. Since 3D models use entire image volumes rather than individual slices, they can exploit more information (e.g. relationships between consecutive slices). Han identified two potential drawbacks of using 3D models: 3D models need more parameters, potentially requiring more training data to achieve robust performance, and 3D models are difficult to implement on commonly-available GPU cards due to their large memory consumption.\cite{RN28} Han therefore used a 2D model rather than a 3D model for brain sCT generation because of the limited available training data and GPU memory\cite{RN28}. Another benefit of using Han`s 2D model\cite{RN28} is that its half weights can be initialized using the pre-trained VGG16 model{\cite{RN29}}. These weights can be used to assist the training process. No such pre-trained weights are available for a 3D model. 

In this paper, we investigated the performance of generating sCTs using CNN models in the male pelvis, which has greater anatomic variation than the brain.  We modified a SegNet-2D CNN model by implementing instance normalization\cite{RN34} and residual shortcuts\cite{RN36} to speed training. We extended the 2D model to 3D to test whether a similar size patient-cohort as in Han\cite{RN28} would be enough to effectively train a 3D model and compared 2D and 3D model performance. We incorporated on-the-fly data augmentation and a modified loss function to enhance model performance. Both models were trained from scratch without implementing transfer learning. Their performance was evaluated and compared using geometric and voxel-wise metrics.

\section{Materials and methods}
\subsection{Image data}  
Retrospective analysis was performed using CT and MR images from 20 prostate cancer patients (61 to 80 years old). The CTs were acquired on a 64-slice CT scanner (Sensation, Siemens Medical Solutions, Erlangen, Germany) using the following settings: 120 kVp, 400 mA, and 1.5 mm or 3 mm slice thickness, with in-plane spatial resolutions varying from 0.85 $\times$ 0.85 mm$^{2}$ to 1.27 $\times$ 1.27 mm$^{2}$. For each patient, an MR image was acquired on the same day as the CT with a non-contrast T1-weighted 2D turbo spin echo sequence (echo time: 12 ms or 13 ms, repetition time: 523 ms to 784 ms, flip angle: 150o) on a 1.5 T MR scanner (Sonata, Siemens Healthcare, Erlangen, Germany). MR images had slice thickness of 5 mm and in-plane spatial resolutions ranging from 0.71 $\times$ 0.71 mm$^{2}$ to 0.94 $\times$ 0.94 mm$^{2}$.  Thirty slices covering the prostate region were extracted from MR images and resampled to dimensions of 256 $\times$ 256 $\times$ 30. The final voxel size of MR images varies from 1.25 $\times$ 1.11 $\times$ 5 mm$^{3}$ to 1.41 $\times$ 1.41 $\times$ 5 mm$^{3}$.

\subsection{Preprocessing}  
\begin{figure}
\begin{center}
\begin{tabular}{c}
\includegraphics[height=12.5cm]{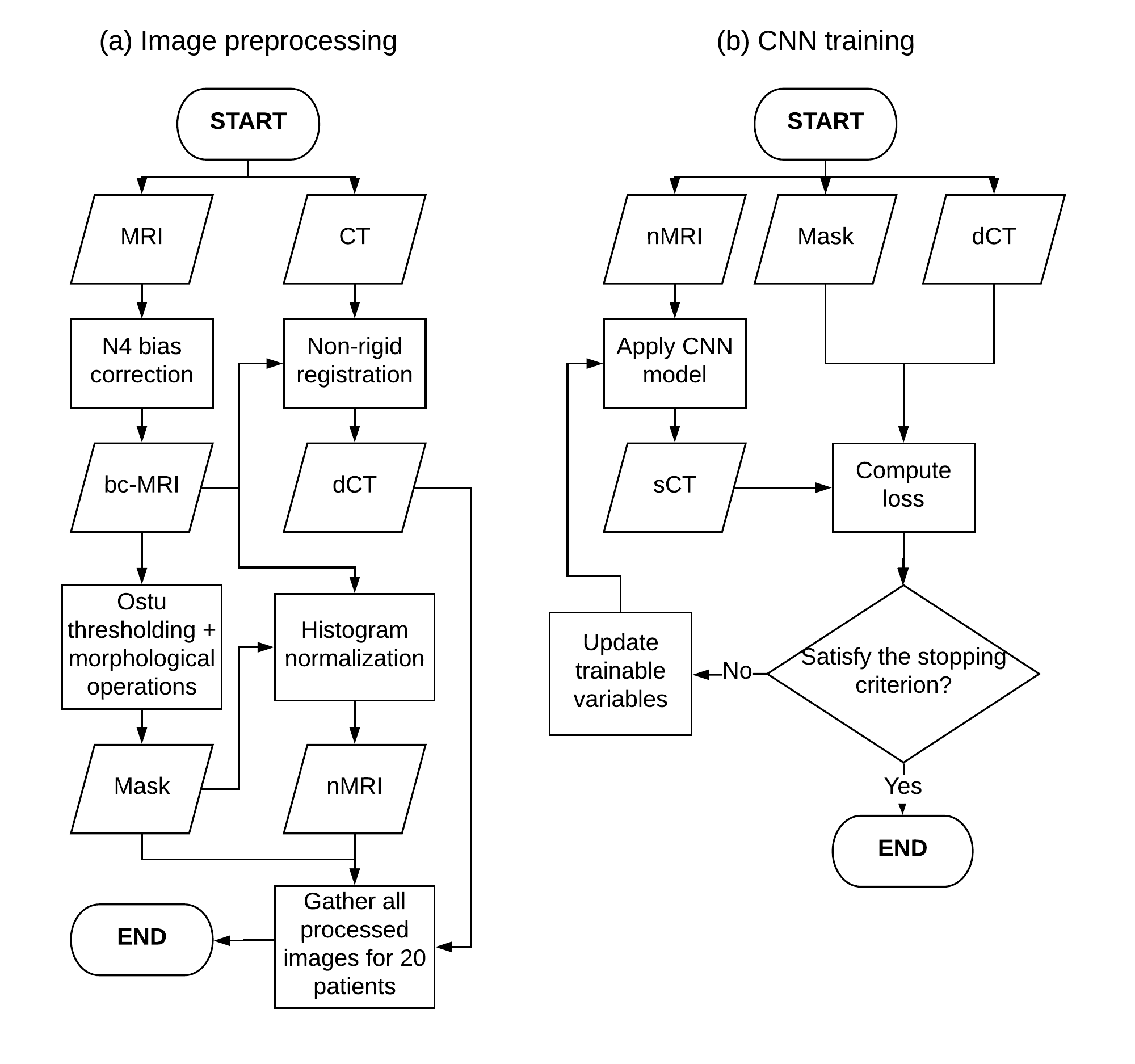}
\end{tabular}
\end{center}
\caption 
{\label{fig1}
The overall workflow of sCT generation. (a) In the preprocessing stage, N4 bias correction was applied to the MRI to get the bias-correct MRI (bc-MRI). The CT was then deformably registered to the bc-MRI to get the paired MRI-deformed CT (dCT). The body mask and normalized MRI (nMRI) were acquired from the bc-MRI for each patient. (b) In the training stage, the sCT was generated by feeding the nMRI into the CNN model. The loss was computed as the mean absolute error between the sCT and dCT within the body mask and then minimized by updating variables of the CNN model using backpropagation and stochastic gradient descent.} 
\end{figure}

Figure~\ref{fig1} outlines the image preprocessing and CNN model training workflows, respectively. N4 bias field correction\cite{RN30} and histogram-based normalization\cite{RN31} were performed on the MR images to minimize the inter-patient intensity variation. A body mask of each patient, which was used for restricting loss evaluation and sCT accuracy assessment, was generated from the bias-corrected MR image using Otsu`s thresholding\cite{RN32} followed by opening and closing morphological operations. To account for organ movement and patient setup variations between CT and MR images, the CT was registered to the bias-corrected MR image using rigid and affine registrations, followed by a multi-resolution B-spline registration (Elastix\cite{RN33}). Each deformed CT (dCT) was resampled to match the MR image resolution. Each dCT was visually compared to its paired MR image to assure that the images were properly registered.

\subsection{2D and 3D CNN models} 
\begin{figure}
\begin{center}
\begin{tabular}{c}
\includegraphics[height=12.5cm]{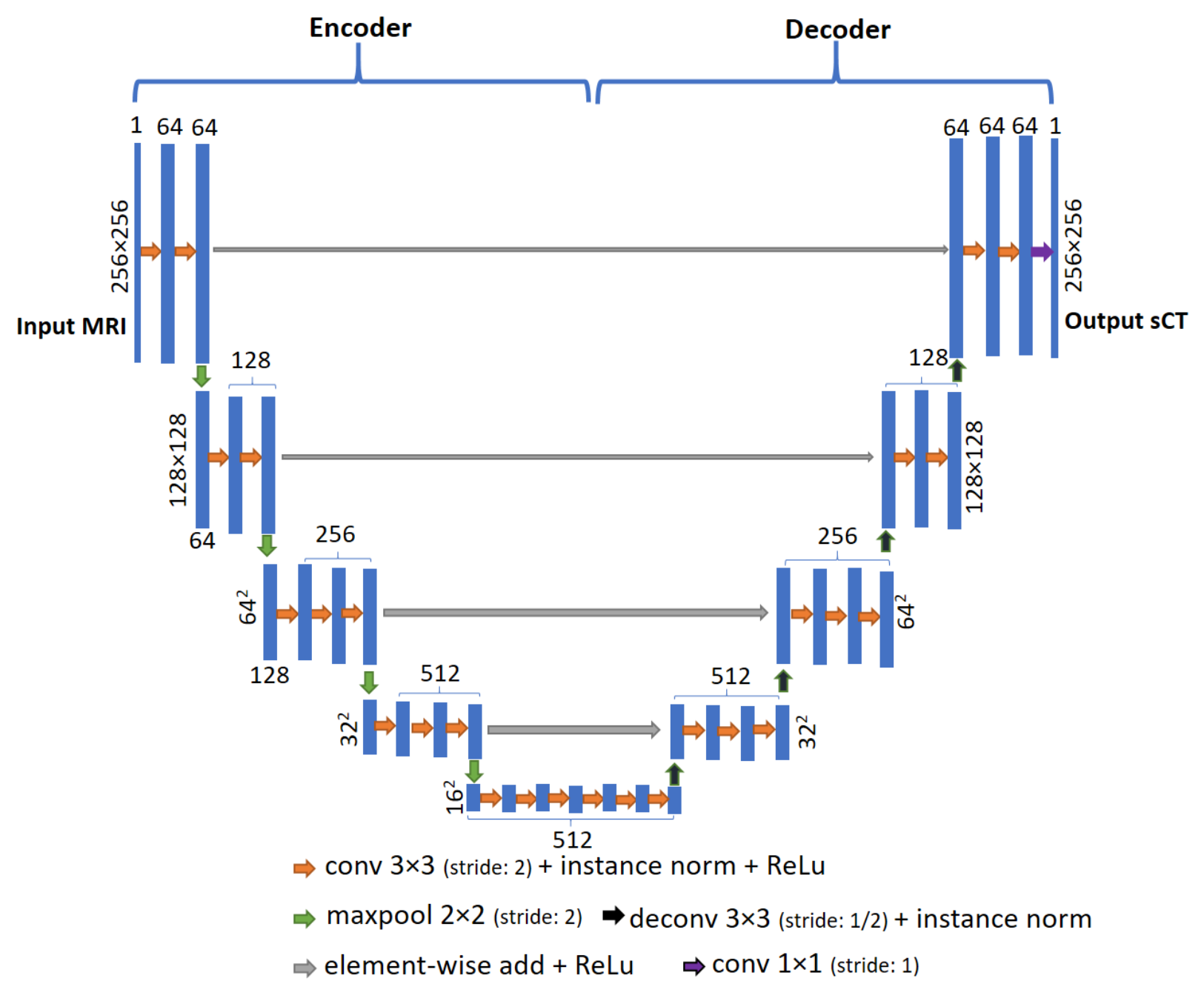}
\end{tabular}
\end{center}
\caption 
{\label{fig2}
The overall 2D CNN model architecture. A slice from the normalized MRI is input into the model. Each blue box represents a set of feature maps whose dimensions and number are shown. Each orange arrow represents a convolutional (conv) layer followed by instance normalization and the activation function (rectified linear unit, ReLu). In the encoder network, a maxpool operation with a 2 $\times$ 2 window and at a stride of 2, shown by green arrows, is applied to reduce the spatial resolution of feature maps, while the deconvolutional (deconv) layer followed by instance normalization layer, shown by black arrows, is used to upsample feature maps. A residual shortcut, shown by gray arrows, is achieved by adding high-resolution feature maps in the encoder network to up-sampled feature maps in the decoder network. Finally, a conv layer consisting of 1×1 filters is used to generate a 2D sCT.} 
\end{figure}

The proposed 2D model was modified from SegNet{\cite{RN27}}, a state-of-the-art deep learning architecture for semantic segmentation, and extended to 3D. 2D MR slices and 3D MR volumes were fed into the corresponding CNN models which were trained to output 2D sCT slices and 3D sCT volumes, respectively. Figure~\ref{fig2} shows the architecture of the 2D model.  

Like SegNet, the 2D model has encoder and decoder networks. The encoder network, consisting of 13 convolutional layers, is identical to the convolutional layers in the VGG16 model{\cite{RN29}}, except that filters in the first convolutional layer have a depth of 1 rather than 3, because of the scalar nature of MR and CT.  Each encoding convolutional layer performed convolution of its input with a set of 3 $\times$ 3 trainable filters at a stride of 1. Zero padding was used to produce feature maps with the same resolution as the inputs. These feature maps were normalized using instance normalization\cite{RN34} to reduce internal covariate shifts and then operated by the element-wise activation function $max(0,x)$, termed the Rectified Linear Unit (ReLU). The feature maps were downsampled by applying a maxpooling layer with a 2 $\times$ 2 window and a stride of 2. The sequence of several convolutional layers and max pooling layers act to extract local and global features and increase translation invariance. 

The decoder network, consisting of a hierarchy of decoders, was used to upsample low-resolution feature maps and gradually reconstruct the sCT. Each decoding convolutional layer corresponded to an encoding convolutional layer, except for the final convolutional layer that had a set of 1 $\times$ 1 learnable filters with a stride of 1.

Three modifications to SegNet\cite{RN27} were made to develop the proposed 2D CNN model. First, the unpooling layers in the original SegNet\cite{RN27} were replaced with fractionally-strided convolutional layers (also known as deconvolutional layers). Unlike unpooling layers, which use memorized pooling indices from maxpooling layers to produce sparse high-resolution feature maps, fractionally-strided convolutional layers can be trained to produce dense high-resolution feature maps.\cite{RN35} Second, residual shortcuts, which element-wise add encoder feature maps to corresponding upsampled feature maps, were introduced for faster convergence. This was inspired by ResNet{\cite{RN36}}. Third, instance normalization\cite{RN34} was employed rather than batch normalization\cite{RN37} to deal with the small batch size. 

The 3D model shared the same architecture as the 2D model except that all 2D operations were replaced with their corresponding 3D counterparts.

The filters in the convolutional layers and fractionally-strided convolutional layers had sets of weights and biases, which were trained by minimizing a loss function. The loss function was defined as the mean absolute error (MAE) between the sCT and deformed CT (dCT) within the body mask;
\begin{equation}
\label{eq1}
loss = \frac{1}{N}\sum_{i=1}^{N}|sCT_{i}-CT_{i}|
\end{equation}
where N was the number of voxels inside the body masks of MR images, and $sCT_{i}$ and $CT_{i}$ represented the HU values of the $i^{th}$ voxel in the sCT and dCT, respectively.

\subsection{Model optimization details}  
Both the 2D and 3D CNN models were implemented using Tensorflow\cite{RN38} packages. The Adam stochastic gradient descent method\cite{RN39} with default parameters, except for the learning rate that was set at 0.01, was used for minimizing the loss function (Equation~\ref{eq1}). At each iteration, a mini-batch of 2D images or 3D volumes was randomly selected from the training set. The batch size was limited by GPU memory. A mini-batch of 15 training slices was used to run the 2D model on an 8 GB NVIDIA GeForce GTX 1080 GPU. The 3D model was run on a 12 GB NVIDIA GeForce GTX Titan X GPU with a mini-batch of 1 training volume. The reduced batch size and large memory GPU card were necessary for implementing the 3D model due to its greater memory consumption. On-the-fly data augmentation (random shift and rotation) was performed on each set of MR images, body masks, and dCTs to reduce overfitting. For both the 2D and 3D models, the random translation was up to 15 pixels in the x and y directions, and the random rotation angle in the x-y plane was confined within $\pm$5$^{o}$. Rotations with random angles within $\pm$2$^{o}$ in the x-z and y-z planes were applied to the 3D images. The 2D and 3D model weights were initialized using He initialization{\cite{RN40}}, and the biases were initialized to 0. 

\subsection{Model evaluation}  
Five-fold-cross-validation was performed to evaluate model performance. The 20 patient-cohort was randomly divided into five groups. Each time validation was performed, four groups were used as the training set to optimize the model. The optimized model was then used to generate sCTs of patients in the remaining group. For the 2D (3D) model, four groups of four patients provided 480 (16) training samples. Using the batch size of 15 (1), it took 32 (16) iterations to go over all samples in the training set for the 2D (3D) model, which was considered as one epoch. 

CNN model accuracy was evaluated by using voxel-wise MAE between the sCT and dCT for three regions: 1) the whole body; 2) a soft tissue region generated by thresholding the dCT with a range [-100,150) HU; and 3) a bone region generated by thresholding the dCT at 150 HU, i.e., [150,$\infty$) HU.

CNN model accuracy was also evaluated by calculating the dice similarity coefficient (DSC), recall, and precision for the bone region. They were defined as:
\begin{equation}
\label{eq2}
DSC = \frac{2(V_{sCT} \cap V_{dCT})}{V_{sCT} + V_{dCT}},
recall = \frac{V_{sCT} \cap V_{dCT}}{V_{dCT}},
precison = \frac{V_{sCT} \cap V_{dCT}}{V_{sCT}}
\end{equation}
where V was the bone-region volume.
Wilcoxon signed-rank tests were performed on the evaluation metrics to test the difference between the performance of 2D and 3D models. $P < 0.05$ was considered statistically significant.

\section{Results}
It required approximately 2 (4) hours to train the 2D (3D) model for 200 epochs using the aforementioned GPU cards. The time required for generating the whole sCT volume of a patient was approximately 5.5 s for both models.  

\begin{figure}
\begin{center}
\begin{tabular}{c}
\includegraphics[height=18cm]{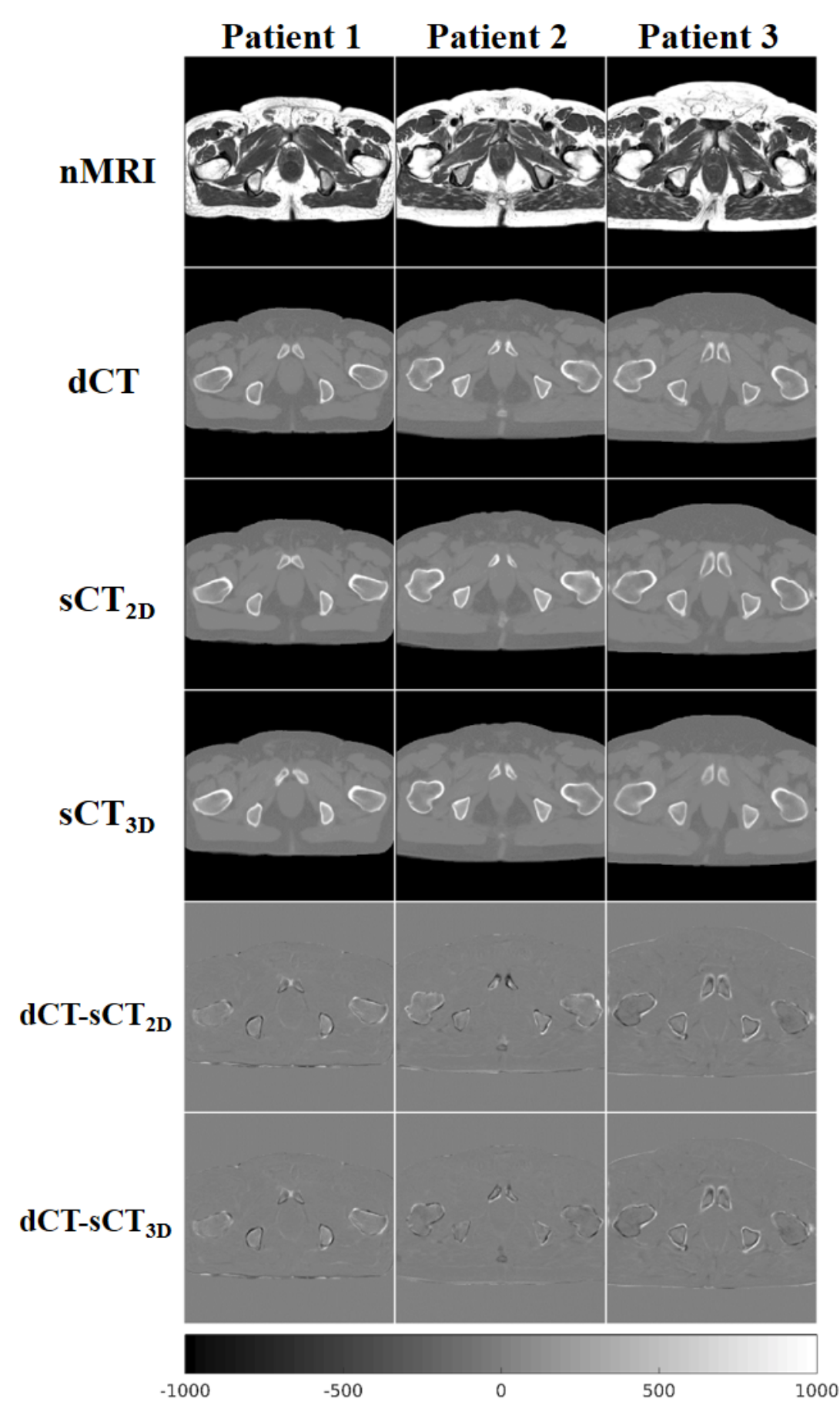}
\end{tabular}
\end{center}
\caption 
{\label{fig3}
Transverse slices of the normalized MRIs (row 1), the dCTs (row 2), the 2D model sCTs (row3) and the 3D model sCTs (row 4) from three patients. The last two rows show the difference maps between the 2D model sCTs and the dCTs (row 5), and the difference maps between the 3D model sCTs and the dCTs (row 6). The color bar is associated with all images except normalized MRIs.} 
\end{figure}

Figure~\ref{fig3} shows transverse slices of sCTs generated by the 2D and 3D models along with the corresponding slices of the normalized T1-weighted MR images and deformed CTs from three patients. As shown in the difference maps, both models gave accurate HU value predictions for most regions, especially soft tissues, but had difficulty generating accurate HU values near the body contour and bone outlines. 

\begin{figure}
\begin{center}
\begin{tabular}{c}
\includegraphics[height=8cm]{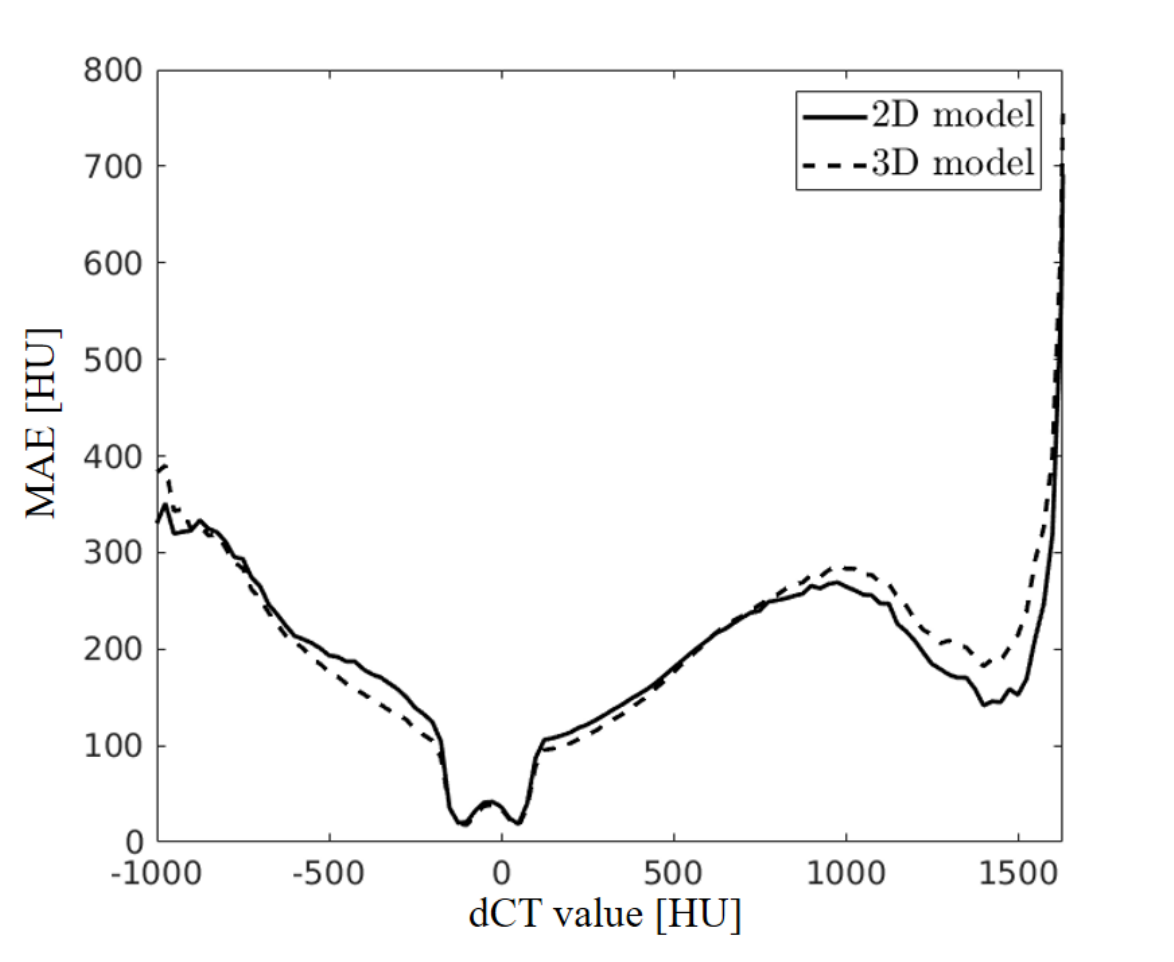}
\end{tabular}
\end{center}
\caption 
{\label{fig4}
MAE of voxels within body masks from all patients as a function of dCT values, calculated in 25 HU bins.} 
\end{figure}

The MAE, including all patients, is shown in Figure~\ref{fig4} as a function of dCT values. The MAE was calculated in 25 HU bins.  Both models behaved similarly, with similar MAE curves for most HU values except that the 2D model yielded greater MAEs than the 3D model within (-650, -200) HU, and vice-versa within (850,1600) HU. 

Table~\ref{tab1} summarizes the statistics of the voxel-wise and geometric metrics averaged across all patients and shows the results of Wilcoxon signed-rank tests. The maximum MAEs within the body were 56.5 HU and 53.1 HU for 2D and 3D models, respectively. The minimum bone-region DSCs were 0.70 and 0.72 for the 2D and 3D models, respectively. As shown in Tab. 1, p-values of all Wilcoxon signed-rank tests were less than 0.05, except for recall. We also performed paired t-tests which yielded the same hypothesis-testing results.

\begin{table}[ht]
\begin{center}
\begin{tabular}{|c|c|c|c|c|}
\hline
\multicolumn{2}{|c|}{} & 2D model & 3D model & P value\\
\hline
\multirow{3}{*}{MAE [HU]} & whole body & 40.5 $\pm$ 5.4 & 37.6 $\pm$ 5.1 & 3.90 $\times$ 10$^{-4}$\\
& soft tissue & 28.9 $\pm$ 4.7 & 26.2 $\pm$ 4.5 & 2.54 $\times$ 10$^{-4}$\\
& bone & 159.7 $\pm$ 22.5 & 154.3 $\pm$ 22.3 & 0.010\\ \hline
\multicolumn{2}{|c|}{DSC} & 0.81 $\pm$ 0.04 & 0.82 $\pm$ 0.04 & 0.048\\
\hline
\multicolumn{2}{|c|}{recall} & 0.85 $\pm$ 0.04 & 0.84 $\pm$ 0.04 & 0.60\\
\hline
\multicolumn{2}{|c|}{precision}  & 0.77 $\pm$ 0.09 & 0.80 $\pm$ 0.08 & 1.7 $\times$ 10$^{-3}$\\
\hline
\end{tabular}
\caption{\label{tab1} Comparison of the MAEs for different HU-thresholded regions, and the DSC, recall, and precision for the bone region for the 2D and 3D models. Results were averaged across the 20-patient cohort. The rightmost column shows the results of Wilcoxon signed-rank tests on the difference of four reported metrics between the two models.} 
\end{center}
\end{table}

\section{Discussion}
In this paper, 2D and 3D CNN models were used to generate pelvic sCTs from T1-weighted MR images. Our sCT generation methods were fully automated, requiring no deformable registration or manual segmentation of bone tissues. As shown in Figure~\ref{fig3}, the 2D and 3D CNN models generated high quality sCTs. MAE curves shown in Figure~\ref{fig4} indicated that both models could precisely estimate soft-tissue HU values but had difficulty in reproducing air and high-density bone tissues. 

The MAEs within the body contour across all patients were 40.5 $\pm$ 5.4 HU and 37.6 $\pm$ 5.1 HU for the 2D and 3D models, respectively. The time required for generating a pelvic sCT using our CNN models was about 5.5 s. Our MAE results are comparable to previous studies. Kim $et \ al.$\cite{RN41} presented a voxel-based weighted summation method that produced an MAE of 74.3 $\pm$ 3.9 HU. However, manual contouring of bone tissues required for this method can be tedious and time-consuming. An MAE of 40.5 $\pm$ 8.2 HU was achieved by Dowling $et \ al.$\cite{RN11} using an average MRI-CT atlas from 38 patients. Andreasen $et \ al.$\cite{RN42} reported an MAE of 54 $\pm$ 8 HU using an atlas-based method with pattern recognition, and its prediction time was about 20.8 min. Another random forest model proposed by Andreasen $et \ al.$\cite{RN43} generated sCTs with an MAE of 58 $pm$ 9 HU. A hybrid method suggested by Siversson $et \ al.$ \cite{RN45} obtained an MAE of 36.5 $\pm$ 4.1 HU when ignoring errors introduced by gas cavities. This hybrid method was implemented in the cloud-based commercial software MriPlanner (Spectronic Medical AB, Helsingborg, Sweden), which required 50 to 80 min to generate a sCT.\cite{RN45} The patch-based 3D context-aware generative adversarial network presented by Nie $et \ al.$\cite{RN26} achieved an MAE of 39.0 $\pm$ 4.6 HU. 

Our CNN models reproduced low-density bone as shown in Figure ~\ref{fig4}. The bone-region DSCs were 0.81 $\pm$ 0.04 and 0.82 $\pm$ 0.04 from the 2D and 3D models, respectively. These results are comparable to reported DSC results of 0.79 $\pm$ 0.12\cite{RN10} and 0.91$\pm$0.03{\cite{RN11}}, where the authors compared bone contours manually drawn on the sCT and CT.

It was feasible to train the proposed 3D model with 16 image volumes from scratch. Results of the Wilcoxon signed-rank tests shown in Table~\ref{tab1} demonstrated a statistically significant improvement in overall MAE, bone DSC, and bone precision of the 3D model compared to the 2D model. However, as shown in Figure~\ref{fig4}, the 2D model seemed to perform better in estimating the high-density bone HU values. It should be noted that smaller overall MAEs do not guarantee improved sCT dose calculation and patient positioning performance. While the models performed well, we will continue to acquire more patient data to potentially improve model accuracy and further test model differences.

As this was a retrospective study, the MR image voxel sizes were not matched, resulting in different voxel intensities between images. This may have affected the sCT generation accuracy although we applied intensity normalization. A potential study could examine how voxel size variations affects sCT estimation. 

The proposed 3D model can be implemented on a 12 GB GPU to process volumetric images with dimensions of 256 $\times$ 256 $\times$ 30. More GPU memory would be required to process higher resolution 3D images. Considering the limited access to multi-GPU systems, a 3D architecture with fewer convolutional layers could be considered to deal with higher resolutions. However, the performance could be affected by the reduced parameters and smaller receptive fields of the less complex model. Another approach would be to extract 30-slice sub-volumes from CT and MR images for training the 3D model. The sCT could then be generated by averaging 30-slice sCT sub-volumes produced by the model. 

A number of techniques could be investigated for improving model performance.  Nie $et \ al.$\cite{RN26} showed that introducing an additional adversarial discriminator improved overall sCT quality. The same approach could be adapted in our proposed 2D and 3D CNN models.  Non-rigid deformation\cite{RN44} could also be applied to both CT and MR images in the process of the on-the-fly data augmentation to produce more training pairs. Multiple MR images acquired with different sequences could be fed into models to provide more information for distinguishing different tissues. Multi-GPU systems with more memory would enable the exploration of larger batch sizes for training CNN models, which could reduce variances in gradient estimation and accelerate the training.

\section{Conclusion}
We presented 2D and 3D CNN models for generating a pelvic sCT from a T1-weighted MR image. In our study, both models successfully generated accurate sCTs for all 20 patients, with a maximum MAEs of 56.5 HU and 53.1 HU for the 2D and 3D models, respectively. Statistical results of 20 patients showed that the 3D model could generate sCTs with better overall MAE, bone DSC, and bone precision. The fast speed and accurate HU mapping of the proposed 2D and 3D CNN models make them promising tools for generating pelvic sCTs for MRI-only radiotherapy. Future work on dose calculation comparisons between the CT and sCT is required before clinical implementation.

\section*{Acknowledgments}
This work was sponsored by Varian Medical Systems. The authors would like to thank Dr. Anand P. Santhanam for the access to GPU clusters.

\bibliographystyle{ama}
\bibliography{ref.bib}   

\end{spacing}
\end{document}